\documentclass{llncs}

\usepackage{amsmath}
\usepackage{amssymb}

\let\emptyset=\varnothing
\newcommand{\lral}{\mathop{\Longrightarrow}\limits}
\newcommand{\ie}{\emph{i.e.}}
\newcommand{\A}{\textup{\bf A}}
\newcommand{\B}{\textup{\bf B}}
\newcommand{\C}{\textup{\bf C}}

\begin{document}

\pagestyle{plain}
\bibliographystyle{plain}

\title{A Generalization of the Lifting Lemma
  for Logic Programming}
\author{ Etienne Payet \and 
         Fred Mesnard
	}
\institute{
	 Iremia - Universit\'e de La R\'eunion, France\\
         \{epayet, fred\}@univ-reunion.fr
	}

\maketitle

\abstract{Since the seminal work of J. A. Robinson on resolution,
many lifting lemmas for simplifying proofs of 
completeness of resolution
have been proposed in the literature.
In the logic programming framework, 
they may also help to detect some infinite derivations
while proving goals under the SLD-resolution. In this paper,
we first generalize a version of the lifting lemma,
by extending the relation
``is more general than'' so that it takes into account
only some arguments of the atoms. The other arguments,
which we call neutral arguments, are  disregarded.
Then we propose two syntactic conditions of increasing
power for identifying neutral arguments from mere inspection
of the text of a logic program. \\
}

\section*{Contents}
\begin{enumerate}
\item Introduction
\item Preliminaries
\item Neutral Arguments for SLD-Derivation
  \begin{enumerate}
  \item[3.1] Filters
  \item[3.2] Derivation Neutral Filters: Operational Definition
  \item[3.3] Model Theoretic Results Induced by DN Filters
  \end{enumerate}
\item Some Particular DN Filters
  \begin{enumerate}
  \item[4.1] DN Sets of Positions
  \item[4.2] DN Sets of Positions with Associated Terms
  \end{enumerate}
\item Atomic Queries with Infinite Derivations
\item Related Works
\item Conclusion
\item Bibliography
\item Proofs
\end{enumerate}

\newpage
\section{Introduction}
Since the seminal work of J. A. Robinson on resolution
\cite{Rob65}, many lifting lemmas 
have been proposed in the literature, see for instance \cite{Chang73} p. 84,
\cite{Apt82a} p. 848,
\cite{Lloyd87a} p. 47, \cite{Apt97a} p. 60 or
\cite{Doe93} where a stronger
version than that of \cite{Apt97a} is given.
Lifting results are used as 
a tool to simplify proofs of completeness of
resolution.

In this paper, we are interested in 
generalizing the lifting lemma presented in \cite{Apt97a}
where it is called the Lifting Theorem.
Given a logic program $P$, our aim is the
further design of a mechanism that generates \emph{at compile-time}
classes of queries that have an infinite SLD-derivation w.r.t. $P$. 
Towards this end, we propose a criterion in the form of a sufficient
condition that enables us to identify infinite derivations.

Notice that such a condition can be easily designed
from the Lifting Theorem that is proposed by Apt in
\cite{Apt97a}, as we explain  now.
Assuming that the reader is familiar with
the standard notations about logic programming reviewed
in the next section,  let us first recall 
the notion of a lift.
An SLD-derivation $\xi'$ is a \emph{lift} of
another SLD-derivation
$\xi$ if there exists a prefix $D$ of $\xi'$ that
is more general than $\xi$ \ie{} $D$ is of the same
length than $\xi$ and
in $D$ and $\xi$
the same clauses are used in the same order, atoms
in the same positions are selected at each step and
for each $i$, the $i$-th query of $D$ is more general
than that of $\xi$. The following result
holds.
\begin{theorem}[Lifting \cite{Apt97a}]
  \label{theorem-classical-lifting}
  Let $P$ be a logic program, $Q$ be a query and
  $\eta$ be a substitution.
  For every SLD-derivation $\xi$
  of $P\cup\{Q\eta\}$,
  there exists an SLD-derivation of
  $P\cup\{Q\}$ which is a lift of $\xi$.
\end{theorem}

This theorem provides a sufficient condition that
can be used to identify infinite SLD-derivations.
\begin{corollary}
  \label{corollary-loop-detection}
  Let $P$ be a logic program and $A$ be an atom.
  Suppose that there exists a sequence of SLD-derivation
  steps from $A$ to a query $Q$ using the clauses of $P$
  and that $Q$ contains an atom $B$ that is more general
  than $A$.
  Then there exists an infinite SLD-derivation of
  $P\cup\{A\}$.
\end{corollary}
\begin{proof}
  Let $A \lral^+_P Q$ denote that
  there exists a sequence of SLD-derivation
  steps from $A$ to $Q$ using the clauses of $P$.
  As $B$ is more general than $A$, by the
  Lifting Theorem~\ref{theorem-classical-lifting} we
  have $B \lral^+_P Q_1$ where
  $Q_1$ is a query that is more general than
  $Q$.
  So there exists $B_1$ in $Q_1$
  that is more general than $B$.
  Iterating this process, we construct an infinite sequence
  of queries $Q_1$, $Q_2$, \dots and an infinite sequence
  of atoms $B_1$, $B_2$, \dots such that for $i\geq 1$,
  $B_i$ is in $Q_i$ and $B_i \lral^+_P Q_{i+1}$.
  \qed
\end{proof}


Nevertheless, such a condition is rather weak because it
fails at identifying some simple loops. This is illustrated
by the following example.

\begin{example}
  \label{example-neutral}
  Let $c$ be the clause
  $p(x,y) \leftarrow p(f(x),z)$. Then from the head
  of $c$ we get an SLD-derivation step
  $p(x,y)\lral_c p(f(x_1),z_1)$.
  Since the atom $p(f(x_1),z_1)$ is not more general than
  $p(x,y)$, Corollary~\ref{corollary-loop-detection} 
  cannot be used. Moreover,
  \[p(f(x_1),z_1) \lral_c p(f(f(x_2)),z_2)\lral_c
    p(f(f(f(x_3))),z_3)\cdots\]
  As the first argument of $p$ grows from step to step,
  we will never
  be able to use Corollary~\ref{corollary-loop-detection}
  to show that there exists an infinite SLD-derivation
  of $\{c\} \cup \{p(x,y)\}$. \qed
\end{example}

In this article, we extend the relation
``is more general than'' so that it takes into account
only some arguments of the atoms, the others
(which are called \emph{neutral arguments}) being,
roughly, disregarded. We show that the
Lifting Theorem~\ref{theorem-classical-lifting}, and
hence Corollary~\ref{corollary-loop-detection},
can be extended to this new relation. Neutral arguments
correspond to the following intuition. Suppose
we have an SLD-derivation $\xi$ of a query $Q$ w.r.t.
a logic program $P$. Let $Q'$ be the query obtained by
replacing the neutral arguments of the atoms of $Q$
by any term. 
Then, there exists an SLD-derivation $\xi'$ of $Q'$
w.r.t. $P$ such that $\xi'$ is a lift of 
$\xi$ up to the neutral arguments.

\begin{example}
Consider Example~\ref{example-neutral} again.
For any derivation step
$p(s_1,s_2)$ $\lral_c p(s_3,s_4)$
if we replace $s_1$ by any term $t_1$
then there exists a derivation step
$p(t_1,s_2) \lral_c p(t_3,t_4)$.
Moreover, notice that $s_4$ and $t_4$ are variables,
so $p(t_3,t_4)$ is more general than $p(s_3,s_4)$
up to the first argument of $p$.
Consequently, by the intuition described above,
the first argument of $p$ is neutral for
derivation w.r.t. $c$.
Finally, as
$p(x,y) \lral_c p(f(x_1),z_1)$ and
$p(f(x_1),z_1)$ is more general than $p(x,y)$ up to the
first argument which is neutral, by
the extended version of
Corollary~\ref{corollary-loop-detection}
there exists an infinite SLD-derivation of
the query $p(x,y)$
w.r.t. $c$. \qed
\end{example}

Last but not least, 
we offer two syntactic conditions of increasing
power for easily identifying neutral arguments from mere inspection
of the text of a logic program.

The paper is organized as follows.
In Section~\ref{sect-prelim}, we
review basic concepts concerning logic programming and
introduce some notations. Then in Section~\ref{sect-neutral}
we give a generic presentation of what we mean for an
argument to be neutral.
In Section~\ref{sect-particular-DN-filters}, we propose
some particular concrete means for detecting neutral arguments.
In Section~\ref{sect-examples}, we
apply our results to generate some queries that have
an infinite SLD-derivation w.r.t. a given logic program.
Finally, in Section~\ref{section-related-works} we
discuss related works.

\section{Preliminaries}
\label{sect-prelim}
We try to strictly adhere to the notations, definitions,
and results presented in \cite{Apt97a}. 

$N$ denotes the set of non-negative integers and for any $n\in N$,
$[1,n]$ denotes the
set $\{1,\ldots,n\}$. If $n=0$ then $[1,n]=\emptyset$.

From now on, we fix a language $\mathcal{L}$ of programs.
We assume that ${\cal L}$ contains an infinite number of
constant symbols.
The set of relation symbols of $\cal L$
is $\Pi$, and we assume that
each relation symbol $p$ has an \emph{unique} arity, 
denoted $arity(p)$. 
$TU_{\cal L}$ (resp. $TB_{\cal L}$)
denotes the set of all (ground and non ground)
terms of ${\cal L}$ (resp. 
atoms of ${\cal L}$). 
A \emph{query} is a finite sequence of atoms
$A_1,\ldots,A_n$ (where $n\geq 0$). Queries are denoted
by $Q$, $Q'$, \dots{} or by bold upper-case letters
$\A$, $\B$, \dots

Let $t$ be a term.
Then $Var(t)$ denotes the set of variables occurring in $t$.
This notation is extended to atoms, queries and clauses.
Let $\theta := \{ x_1/t_1,\ldots,x_n/t_n \}$ be a substitution. 
We denote by $Dom(\theta)$ the set of variables 
$\{x_1,\ldots,x_n\}$, and by $Ran(\theta)$ the set
of variables appearing in $t_1,\ldots,t_n$. We define 
$Var(\theta)=Dom(\theta) \cup Ran(\theta)$. Given a set of
variables $V$, $\theta|V$ denotes the substitution
obtained from $\theta$ by restricting its domain to $V$.

Let $t$ be a term and $\theta$ be a substitution. Then,
the term $t\theta$ is called an \emph{instance} of $t$.
If $\theta$ is a \emph{renaming} (\ie{} a substitution that is
a 1-1 and onto mapping from its domain to itself),
then $t\theta$ is called a
\emph{variant} of $t$. Finally, $t$ is called
\emph{more general than $t'$} if $t'$ is an instance of
$t$.

A \emph{logic program} is a finite set of definite clauses. 
In program examples, we use the ISO-Prolog syntax.
Let $P$ be a logic program. Then $\Pi_{P}$ denotes the set
of relation symbols appearing in $P$.
Consider a non-empty query $\A,B,\C$
and a clause $c$.
Let $H\leftarrow \mathbf{B}$ be a variant of $c$
variable disjoint
with $\A,B,\C$ and
assume that $B$ and $H$ unify. Let $\theta$ be an mgu
of $B$ and $H$. Then 
$\A,B,\C \lral_{c}^{\theta}  
(\A,\mathbf{B},\C)\theta$ 
is an \emph{SLD-derivation step} with $H \leftarrow \mathbf{B}$
as its \emph{input clause} and $B$ as its \emph{selected atom}.
If the substitution $\theta$ or the clause $c$ is irrelevant,
we drop a reference to it.

Let $Q_0$ be a query. A maximal sequence
$Q_0 \lral_{c_1}^{\theta_1} Q_1 \lral_{c_2}^{\theta_2} \cdots$
of SLD-derivation steps is called an \emph{SLD-derivation}
of $P\cup\{Q_0\}$ if $c_1$, $c_2$, \dots are clauses of $P$
and if the \emph{standardization apart} condition holds, \ie:
each input clause used is variable disjoint from the
initial query $Q_0$ and from the mgu's and input clauses
used at earlier steps.
A finite SLD-derivation may end up either 
with the empty query (then it is a \emph{successful}
SLD-derivation) or with a non-empty query
(then it is a \emph{failed} SLD-derivation). 
We say $Q_0$ \emph{loops} with respect to $P$
if there exists an infinite SLD-derivation of $P\cup\{Q_0\}$.

Finally, we write $Q \lral_{P}^{*} Q'$
(resp. $Q \lral_{P}^{+} Q'$) if there exists a finite
prefix (resp. a finite non-empty prefix)
ending at $Q'$ of an SLD-derivation of $P\cup\{Q\}$.

Let $E$ and $F$ be two sets. Then, $f : E \rightarrow F$ denotes
that $f$ is a function from $E$ to $F$ and
$f : E \rightarrowtail F$ denotes
that $f$ is a mapping from $E$ to $F$. The \emph{domain} of a
function $f$ from $E$ to $F$ is denoted by $Dom(f)$ and is
defined as:
$Dom(f) = \{x \; | \; x\in E, \; f(x) \text{ exists}\}$.
Thus, if $f$ is a mapping from $E$ to $F$, then $Dom(f)=E$.

\section{Neutral Arguments for SLD-Derivation}
\label{sect-neutral}

The basic idea in the work we present here relies on
some arguments in clauses which we can be disregarded
when unfolding a query. For instance, the second argument
of the non-unit clause of the program
\begin{verbatim}
    append([],Ys,Ys).
    append([X|Xs],Ys,[X|Zs]) :- append(Xs,Ys,Zs).
\end{verbatim}

is such a candidate. 
Notice that a very common programming technique called
\emph{accumulator passing} (see for instance  e.g.
\cite{Okeefe90a}, p. 21--25),
always produces such patterns.
\begin{example}\label{ex-rev}
  A classical example of the accumulator passing
  technique is the following program \texttt{REVERSE}.
  \begin{verbatim}
    reverse(L,R) :- rev(L,[],R).
    rev([],R,R).
    rev([X|Xs],R0,R) :- rev(Xs,[X|R0],R).
  \end{verbatim}
  Concerning termination, we may ignore the third
  \emph{and} the second argument of the recursive
  clause of \texttt{rev} while unfolding a query with
  this clause. Only the first argument can stop the
  unfolding.
  \qed
\end{example}

But we can be more precise. Instead of
only identifying arguments that can be totaly
disregarded as in the above examples, we can try
to identify arguments that can be replaced,
when unfolding a query, by any terms for which
a given condition holds.
Consider for instance the program
\begin{verbatim}
    p(x,f(z)) :- q(x,y), p(y,f(z)).
\end{verbatim}
If we unfold a query $p(t_1,t_2)$ with this
program, then if we replace the second argument
of $p$ by any instance $t_3$ of $f(z)$, we can
still unfold $p(t_1,t_3)$.

In the sequel of this section, we give a technical
tool to describe specific arguments inside
a program and present an extension of
the relation ``is more general than''.
Then we formalize
the concept of derivation neutrality and we
propose an extended version of the
Lifting Theorem~\ref{theorem-classical-lifting}.

\subsection{Filters}
Let us first introduce the notion of a \emph{filter}.
We use filters in order to distinguish atoms some arguments
of which satisfy a given condition. A condition upon
atom arguments, \ie{} terms,
can be defined as a mapping in the following way.

\begin{definition}[Term-Condition]
  A \emph{term-condition} is a mapping from $TU_{\mathcal{L}}$ to
  $\{\mathtt{true},\mathtt{false}\}$.
\end{definition}
\begin{example}\label{example-term-condition}
  The following mappings are term-conditions.
  \[\begin{array}{rccl}
     f_{\mathit{true}} : & TU_{\mathcal{L}} & \rightarrowtail &
     \{\texttt{true}, \texttt{false}\} \\
     & t & \mapsto & \texttt{true} \\[2ex]
     f_1 : & TU_{\mathcal{L}} & \rightarrowtail &
     \{\texttt{true}, \texttt{false}\}\\[1ex]
     & t & \mapsto & \texttt{true}
     \text{ iff $t$ is an instance of $[x|y]$}\\[2ex]
     f_2 : & TU_{\mathcal{L}} & \rightarrowtail &
     \{\texttt{true}, \texttt{false}\} \\[1ex]
     & t & \mapsto & \texttt{true}
     \text{ iff $t$ unifies with $h(a,x)$}
  \end{array}\]
\end{example}

Now we can precise what we exactly mean by a filter.
\begin{definition}[Filter]
  A \emph{filter}, denoted by $\Delta$, is a mapping
  from $\Pi$ such that: for each $p \in \Pi$,
  $\Delta(p)$ is a function from $[1,arity(p)]$ to
  the set of term-conditions.
\end{definition}
\begin{example}
  [Example~\ref{example-term-condition} continued]%
  \label{example-criterion}
  Let $\Pi:=\{p\}$ with $p$ a relation symbol whose arity equals 3.
  Then,
  $\Delta:=\langle\;p\mapsto\langle 1\mapsto f_{\mathit{true}},
   \;2\mapsto f_1\rangle \;\rangle$
  is a filter.
  \qed
\end{example}

Notice that, given a filter $\Delta$,  the relation
``is more general than'' can be extended in the following way:
an atom $A:=p(\cdots)$ is $\Delta$-more general than
$B:=p(\cdots)$ if
the ``is more general than'' requirement holds
for those arguments of $A$ whose position
is not in the domain of $\Delta(p)$ while the other arguments
satisfy their associated term-condition. 
\begin{definition}[$\Delta$-More General]
  \label{definition-Delta-subsumes}
  Let $\Delta$ be a filter, $A$ and $B$ be two atoms and
  $Q:=A_1,\dots,A_n$ and $Q':=B_1,\dots,B_m$ be two
  queries.
  \begin{itemize}
  \item Let $\eta$ be a substitution. Then
    \emph{$A$ is $\Delta$-more general than $B$ for $\eta$} if:
    \[\left\lbrace
    \begin{array}{l}
      A=p(s_1,\dots,s_n)\\
      B=p(t_1,\dots,t_n)\\
      \forall i \in [1,n]\setminus Dom(\Delta(p)), \; t_i = s_i\eta\\
      \forall i \in Dom(\Delta(p)), \; \Delta(p)(i)(s_i) = \mathtt{true}\,.
    \end{array}\right.\]
  \item \emph{$A$ is $\Delta$-more general than $B$} if
    there exists a substitution $\eta$ s.t. $A$ is 
    $\Delta$-more general than $B$ for $\eta$.
  \item Let $\eta$ be a substitution. Then
    \emph{$Q$ is $\Delta$-more general than $Q'$ for $\eta$} if:
    \[\left\lbrace
    \begin{array}{l}
      n=m \text{ and}\\
      \forall i \in [1,n], \; A_i
      \text{ is $\Delta$-more general than } B_i
      \text{ for } \eta\;.
    \end{array}\right.\]
  \item \emph{$Q$ is $\Delta$-more general than $Q'$} if
    there exists a substitution $\eta$ s.t. $Q$ is
    $\Delta$-more general than $Q'$ for $\eta$.
  \end{itemize}
\end{definition}
\begin{example}%
  \label{example-f-subsumes}
  Let $\Pi:=\{p\}$ and $A := p(b,x,h(a,x))$, $B:=p(a,[a|b],x)$,
  $C:=p(a,[a|b],h(y,b))$.
  \begin{itemize}
  \item Consider the filter $\Delta$ defined in
    Example~\ref{example-criterion}. Then,
    $A$ is not $\Delta$-more general than $B$ and $C$ because,
    for instance, its second argument $x$
    is not an instance of $[x|y]$ as required by $f_1$.
    On the other hand, $B$ is $\Delta$-more general than
    $A$ for the substitution $\{x/h(a,x)\}$ and $B$ is
    $\Delta$-more general than $C$ for the substitution
    $\{x/h(y,b)\}$.
    Finally, $C$ is not $\Delta$-more general than $A$ because
    $h(y,b)$ is not more general than $h(a,x)$ and $C$ is not
    $\Delta$-more general than $B$ because $h(y,b)$ is not more
    general than $x$.
  \item Consider the term-conditions defined in
    Example~\ref{example-term-condition}. Let
    $\Delta':=\langle\;p\mapsto\langle 1\mapsto f_{\mathit{true}},
     \;2\mapsto f_1,\;3\mapsto f_2 \rangle \;\rangle$.
    Then, $A$ is not $\Delta'$-more general than $B$ and $C$ for the
    same reason as
    above. On the other hand, $B$ is $\Delta'$-more general than
    $A$ and $C$ for any substitution and $C$ is
    $\Delta'$-more general than $A$ and $B$ for any substitution.
    \qed
  \end{itemize}
\end{example}

The following proposition states an intuitive result: 
\begin{proposition}\label{prop-var-included}
  Let $\Delta$ be a filter and $Q$ and $Q'$ be two queries.
  Then $Q$ is $\Delta$-more general than $Q'$ if and only if
  there exists a substitution $\eta$ such that
  $Var(\eta) \subseteq Var(Q,Q')$ and $Q$ is
  $\Delta$-more general than $Q'$ for $\eta$.
\end{proposition}
\begin{proof}
  The proof of this proposition is given in Appendix 1.
  \qed
\end{proof}

\subsection{Derivation Neutral Filters:
  Operational Definition}
Before we give a precise definition of the kind of
filters we are interested in, we review the notion
of a \emph{lift}.
The definition we propose below is the same as
that of \cite{Apt97a} p. 57 up to the arguments
whose position is distinguished by a filter.
\begin{definition}[$\Delta$-Lift]
  Let $\Delta$ be a filter. Consider a sequence of
  SLD-derivation steps 
  \[\xi:=Q_0 \lral_{c_1} Q_1 \cdots Q_n \lral_{c_{n+1}} Q_{n+1}
    \cdots\]
  We say that the sequence of SLD-derivation steps 
  \[\xi':=Q'_0 \lral_{c_1} Q'_1 \cdots Q'_n \lral_{c_{n+1}} Q'_{n+1}
    \cdots\]
  is a \emph{$\Delta$-lift} of $\xi$ if
  \begin{itemize}
  \item $\xi$ is of the same or smaller length than $\xi'$,
  \item for each $Q_i$ in $\xi$, $Q'_i$ is
    $\Delta$-more general than $Q_i$,
  \item for each $Q_i$ in $\xi$, in $Q_i$ and $Q'_i$ atoms
    in the same positions are selected.
  \end{itemize}
\end{definition}

In the sequel of this paper, we focus on
``derivation neutral'' filters.
The name ``derivation neutral'' stems from the fact
that in any derivation of a query $Q$, the arguments 
of $Q$ whose position is distinguished by such a
filter can be safely replaced 
by any terms satisfying the associated term-condition.
Such a replacement does not modify the derivation process.
\begin{definition}[Derivation Neutral]
  \label{definition-derivation-neutral}
  Let $\Delta$ be a filter and $c$ be a clause.
  We say that \emph{$\Delta$ is DN for $c$} if
  \begin{itemize}
  \item for each SLD-derivation step $Q\lral_c Q_1$,
  \item for each query $Q'$ that is 
    $\Delta$-more general than $Q$,
  \end{itemize}
  there exists a query $Q'_1$ such that $Q'\lral_c Q'_1$
  and $Q'\lral_c Q'_1$ is a $\Delta$-lift of $Q\lral_c Q_1$.

  This definition is extended to finite sets of clauses:
  \emph{$\Delta$ is DN for a logic program $P$} if it is
  DN for  each clause of $P$.
\end{definition}
\begin{example}
  The following examples illustrate the above definitions.
  \begin{itemize}
  \item Consider the following program \texttt{APPEND}:
    \begin{verbatim}
      append([],Xs,Xs).                              % C1
      append([X|Xs],Ys,[X|Zs]) :- append(Xs,Ys,Zs).  % C2
    \end{verbatim}
    Consider the term-condition $f_{\mathit{true}}$ defined in
    Example~\ref{example-term-condition}, the set of relation symbols
    $\Pi:=\{append\}$ and the filter
    $\Delta:=\langle append \mapsto \langle 2 \mapsto
     f_{\mathit{true}} \rangle \rangle$.
    Then, $\Delta$ is DN for \verb+C2+. However, $\Delta$ is not DN for
    \texttt{APPEND} because it is not DN for \verb+C1+.
  \item Consider the following program \texttt{MERGE}:
    \begin{center}
      \rm \texttt{merge([X|Xs],[Y|Ys],[X|Zs]) :- merge(Xs,[Y|Ys],Zs).}
    \end{center}
    Consider the term-condition $f_1$ defined in
    Example~\ref{example-term-condition}, the set of relation symbols
    $\Pi:=\{merge\}$ and the filter
    $\Delta:=\langle merge \mapsto \langle2 \mapsto f_1 \rangle \rangle$.
    Then, $\Delta$ is DN for \texttt{MERGE}. \qed
  \end{itemize}
\end{example}

Repeatedly using this definition, we get an extended version
of the Lifting Theorem~\ref{theorem-classical-lifting}.
\begin{theorem}[$\Delta$-Lifting]
  \label{lifting-theorem}
  Let $P$ be a logic program and $\Delta$ be a DN filter
  for $P$.
  Let $\xi$ be an SLD-derivation of $P\cup\{Q_0\}$
  and $Q'_0$ be a query that is
  $\Delta$-more general than $Q_0$.

  Then there exists an SLD-derivation of $P\cup\{Q'_0\}$
  that is a $\Delta$-lift of $\xi$.
\end{theorem}
\begin{proof}
  If $\xi$ is infinite, we can construct a required
  SLD-derivation $\xi'$ whose length is infinite
  by repeatedly using
  Definition~\ref{definition-derivation-neutral}.

  Now if $\xi$ is of finite length, then
  \begin{itemize}
  \item either $\xi$ is successful: in this case, we can construct
    a required SLD-derivation $\xi'$ of the same length than $\xi$
    by repeatedly using
    Definition~\ref{definition-derivation-neutral},
  \item either $\xi$ fails. Suppose that
    $\xi := Q_0 \lral_{c_1} Q_1 \cdots Q_n \lral_{c_{n+1}} Q_{n+1}$.
    Then, by repeatedly using
    Definition~\ref{definition-derivation-neutral},
    we can construct
    a finite sequence of SLD-derivation steps
    \[Q'_0 \lral_{c_1} Q'_1 \cdots Q'_n \lral_{c_{n+1}} Q'_{n+1}\]
    that is a $\Delta$-lift of $\xi$.
    But as $Q'_{n+1}$ is $\Delta$-more general than $Q_{n+1}$,
    this sequence may not be a maximal one, \ie{} from
    $Q'_{n+1}$ and $P$, we may derive
    $Q'_{n+2}$ and so forth. Hence, in this case, there
    may exist an SLD-derivation of $P\cup\{Q'_0\}$
    that is a $\Delta$-lift of $\xi$ but of longer length
    than $\xi$.
    \qed
  \end{itemize}
\end{proof}

\subsection{Model Theoretic Results
  Induced by DN Filters}
Lifting lemmas are used in the literature to prove
completeness of SLD-resolution.
Now that we have established an extended Lifting Theorem,
it may be worth to investigate its consequences
from the model theoretic point of view.

Let $E$ be an atom or a query
and $\Delta$ be a filter.
Then one may ``expand'' the atoms occurring in $E$ by replacing
every argument whose position is distinguished by $\Delta$ by any
term that satisfies the associated term-condition.
\begin{definition}[Expansion of atoms and queries by a filter]
  Let $\Delta$ be a filter.
  \begin{itemize}
  \item Let $A$ be an atom.
    \emph{The expansion of $A$ w.r.t. $\Delta$},
    denoted $[A]^{\Delta}$, is the set
    defined as
    \[[A]^{\Delta} := \{B \in TB_{\mathcal{L}} \; | \;
      B \text{ is $\Delta$-more general than $A$ for }
      \epsilon\}\]
      where $\epsilon$ denotes the empty substitution.
  \item Let $Q := A_1,\dots,A_n$ be a query.
    \emph{The expansion of $Q$ w.r.t. $\Delta$},
    denoted $[Q]^{\Delta}$, is the set defined as:
    \[[Q]^{\Delta} := \{ (B_1,\dots,B_n) \;|\;
      B_1 \in [A_1]^{\Delta}, \dots, B_n\in[A_n]^{\Delta}
    \}.\]
  \end{itemize}
\end{definition}

Term interpretations in the context of logic programming
were introduced in~\cite{clark79a} and further
investigated in~\cite{Deransart87}
and~\cite{Falaschi93}.
A term interpretation for $\mathcal{L}$ is
identified with a (possibly empty) subset of the
term base $TB_{\mathcal{L}}$. So, as for
atoms, a term interpretation can be
expanded by a set of positions.
\begin{definition}[Expansion of term interpretations
    by a filter]
  Let $\Delta$ be a filter and
  $I$ be a term interpretation for $\mathcal{L}$.
  Then $[I]^{\Delta}$ is the term interpretation for
  $\mathcal{L}$ defined as:
  \[[I]^{\Delta} := \bigcup_{A \in I} [A]^{\Delta}\;.\]
\end{definition}

For any logic program $P$, we denote by $\mathcal{C}(P)$
its least term model.
\begin{theorem}
  Let $P$ be a logic program and $\Delta$ be a DN filter 
  for $P$. Then
  $[\mathcal{C}(P)]^{\Delta} = \mathcal{C}(P)$.
\end{theorem}
\begin{proof}
  The inclusion
  $\mathcal{C}(P) \subseteq [\mathcal{C}(P)]^{\Delta}$
  is trivial so let us concentrate on the other one \ie{}
  $[\mathcal{C}(P)]^{\Delta}\subseteq\mathcal{C}(P)$.
  Let $A' \in [\mathcal{C}(P)]^{\Delta}$. Then there
  exists $A \in \mathcal{C}(P)$ such that
  $A'\in [A]^{\Delta}$. A well known result states:
  \begin{equation}\label{theorem-CP-eq1}
    \mathcal{C}(P) = \{B\in TB_{\mathcal{L}} \; | \;
    \text{there exists a successful SLD-derivation of } B\}
  \end{equation}
  Consequently, there exists a successful SLD-derivation of
  $A$. Therefore, by the $\Delta$-Lifting
  Theorem~\ref{lifting-theorem}, there exists a successful
  SLD-derivation of $A'$. So by~(\ref{theorem-CP-eq1})
  $A'\in\mathcal{C}(P)$.
  \qed
\end{proof}

\section{Some Particular DN Filters}
\label{sect-particular-DN-filters}
In this section, we consider some instances of
the definitions of Section~\ref{sect-neutral}.

\subsection{DN Sets of Positions}
\label{subsection-DN-sets-of-positions}
The first instance we consider 
corresponds to filters whose associated
term-conditions are all equal to $f_{\mathit{true}}$  
(see Example~\ref{example-term-condition}.)
Within such a context, as the term-conditions are fixed,
each filter $\Delta$ is uniquely determined by
the domains of the functions $\Delta(p)$
for  $p\in\Pi$.
Hence the following definition.

\begin{definition}[Set of Positions]
  \label{definition-set-of-positions}
  A \emph{set of positions}, denoted by $\tau$, is a
  mapping from $\Pi$ to $2^N$ such that:
  for each $p\in\Pi$, $\tau(p)$ is a subset of
  $[1,arity(p)]$.
\end{definition}
\begin{example}\label{example-set-of-pos}
  Let $\Pi := \{\mathit{append}, \mathit{append3}\}$
  with $\mathit{arity}(\mathit{append})=3$ and
  $\mathit{arity}(\mathit{append3})=4$. Then
  \[\tau := \langle \;\mathit{append} \mapsto \{2\},
    \; \mathit{append3} \mapsto \{2,3\}\;\rangle\]
  is a set of positions. \qed
\end{example}

Not surprisingly, the filter that is generated by a
set of positions is defined as follows.
\begin{definition}[Associated Filter]
  \label{definition-set-of-pos-associated-filter}
  Let $\tau$ be a set of positions
  and $f_{\mathit{true}}$ be the term-condition defined in
  Example~\ref{example-term-condition}.
  The filter $\Delta[\tau]$ defined as:
  \[\textrm{for each } p\in \Pi, \; \Delta[\tau](p)
    \textrm{ is the mapping from }
    \tau(p) \textrm{ to } \{f_{\mathit{true}}\}
  \]
  is called \emph{the filter associated to $\tau$}.
\end{definition}
\begin{example}[Example~\ref{example-set-of-pos} continued]
  The filter associated to $\tau$ is
  \[\Delta[\tau] := \langle \;\mathit{append} \mapsto
    \langle 2 \mapsto f_{\mathit{true}}\rangle, \; 
    \mathit{append3} \mapsto \langle 2 \mapsto f_{\mathit{true}},\;
    3 \mapsto f_{\mathit{true}} \rangle \;\rangle\;.\]
\end{example}

Now we define a particular
kind of sets of positions. These are named after
``DN'' because, as stated by
Theorem~\ref{theorem-set-of-positions} below,
they generate DN filters.
\begin{definition}[DN Set of Positions]
  \label{definition-DN-set-of-positions}
  Let $\tau$ be a set of positions. We say that
  \emph{$\tau$ is DN for a clause}
  $p(s_1,\dots,s_n)$ $\leftarrow \B$ if:
  \[\forall i \in \tau(p), \;
  \left\lbrace
    \begin{array}{l}
      s_i \text{ is a variable}\\
      s_i \text{ occurs only once in } p(s_1,\dots,s_n)\\
      \text{for each } q(t_1,\dots,t_m) \in \B: \\
      \hspace{0.5cm}\forall j \in [1,m],\;  s_i \in Var(t_j)
       \Rightarrow j \in \tau(q)\;.
    \end{array}\right.\]
A set of positions is \emph{DN for a  program $P$}
if it is DN for each clause of $P$.
\end{definition}

The intuition of
Definition~\ref{definition-DN-set-of-positions}
is the following. If for instance we have a clause
$c:=p(x,y,f(z))\leftarrow p(g(y,z),x,z)$
then in the first two positions of $p$ we can put
any terms and get a derivation step w.r.t. $c$ because
the first two arguments of the head of $c$ are variables
that appear exactly once in the head. Moreover,
$x$ and $y$ of the head reappear in the body but again only
in the first two positions of $p$. So, if we have a derivation
step $p(s_1,s_2,s_3) \lral_c p(t_1,t_2,t_3)$, we can
replace $s_1$ and $s_2$ by any terms $s'_1$ and $s'_2$
and get another derivation step
$p(s'_1,s'_2,s_3) \lral_c p(t'_1,t'_2,t'_3)$ where
$t'_3$ is the same as $t_3$ up to variable names.

\begin{example}[Example~\ref{example-set-of-pos} continued]
  $\tau$ is DN for the following program:
  \begin{verbatim}
  append([X|Xs],Ys,[X|Zs]) :- append(Xs,Ys,Zs).
  append3(Xs,Ys,Zs,Ts) :- append(Xs,Ys,Us), append(Us,Zs,Ts).
  \end{verbatim}
\end{example}

DN sets of positions generate DN filters.
\begin{theorem}\label{theorem-set-of-positions}
  Let $\tau$ be a DN set of positions for a logic program $P$.
  Then $\Delta[\tau]$ is DN for $P$.
\end{theorem}
\begin{proof}
  See Lemma~\ref{prop-special-case} and
  Theorem~\ref{theorem-DN-associated}
  further.
  \qed
\end{proof}

Notice that the set of DN sets of positions of a 
logic program is not empty as stated by the following
proposition.
\begin{proposition}
  Let $P$ be a logic program. Then
  $\tau_0 := \langle p \mapsto \emptyset \; | \; 
   p \in \Pi\rangle$
  is DN for $P$.
\end{proposition}
\begin{proof}
  By Definition~\ref{definition-DN-set-of-positions}.
  \qed
\end{proof}
Moreover, an atom $A$ is $\Delta[\tau_0]$-more general
than an atom $B$ iff $A$ is more general than $B$.
So, in the context of the filter $\Delta[\tau_0]$,
the $\Delta$-Lifting Theorem~\ref{lifting-theorem}
is the same as the
Lifting Theorem~\ref{theorem-classical-lifting}.

\subsection{DN Sets of Positions with Associated Terms}
\label{subsection-DN-sets-with-associated-terms}

Now we consider another instance of the definitions
of Section~\ref{sect-neutral}. As we will see, it
is more general than the previous one.
It corresponds to filters whose associated
term-conditions have all the form
``is an instance of $t$'' where $t$ is a term that
uniquely determines the term-condition.
Hence the following definition.
\begin{definition}[Sets of Positions with Associated Terms]
  A \emph{set of positions with associated terms}, denoted
  by $\tau^+$, is a mapping from $\Pi$ such that:
  for each $p\in\Pi$, $\tau^+(p)$ is a function from
  $[1,arity(p)]$ to $TU_{\mathcal{L}}$.
\end{definition}
\begin{example}\label{example-set-of-pos-with-associated}
  Let $\Pi:=\{p,q\}$ where the arity of $p$ and $q$ is 2.
  Then,
  \[\tau^+:=\langle \; p \mapsto
    \langle 2 \mapsto x\rangle,\;
    q\mapsto \langle 2\mapsto g(x)\rangle\;\rangle\]
  is a set of positions with associated terms. \qed
\end{example}

The filter that is generated by a
set of positions with associated terms is defined as follows.
\begin{definition}[Associated Filter]
  \label{definition-associated-filter}
  Let $\tau^+$ be a set of positions with associated terms.
  The \emph{filter associated to $\tau^+$},
  denoted by $\Delta[\tau^+]$, is defined as: for each
  $p\in\Pi$, $\Delta[\tau^+](p)$ is the mapping
    \[\begin{array}{ccl}
        Dom(\tau^+(p)) & \rightarrowtail &
	\text{The set of term-conditions}\\[1ex]
	i & \mapsto & \left\lbrace
	                \begin{array}{ccl}
			  TU_{\mathcal{L}} & \rightarrowtail &
			  \{\mathtt{true},\mathtt{false}\}\\[1ex]
			  t & \mapsto & \mathtt{true}
			  \text{ iff $t$ is an instance of } \tau^+(p)(i)\\
			\end{array}\right.
    \end{array}\]
\end{definition}
\begin{example}
  [Example~\ref{example-set-of-pos-with-associated} continued]
  The filter associated to $\tau^+$ is
  \[\Delta[\tau^+]:=\langle\; p \mapsto \langle 2 \mapsto f_1\rangle,\;
    q \mapsto \langle 2 \mapsto f_2\rangle\;\rangle \text{ where}\]
    \[\begin{array}{rccl}
      f_1 : & TU_{\mathcal{L}} & \rightarrowtail &
	\{\texttt{true},\texttt{false}\}\\[1ex]
      & t & \mapsto & \texttt{true} 
	\text{ iff $t$ is an instance of } x\\[4ex]
      f_2 : & TU_{\mathcal{L}} & \rightarrowtail & 
        \{\texttt{true},\texttt{false}\}\\[1ex]
      & t & \mapsto & 
	\texttt{true} \text{ iff $t$ is an instance of } g(x)		    
    \end{array}\]
\end{example}

As for sets of positions, we define a special kind
of sets of positions with associated terms.
\begin{definition}[DN Sets of Positions with Associated Terms]
  \label{def-DN}
  Let $\tau^+$ be a set of positions with associated terms.
  We say that $\tau^+$ is \emph{DN for a clause}
  $p(s_1,\dots,s_n) \leftarrow \B$ if these conditions hold:
  \begin{itemize}
  \item \textrm{\bf(DN1)}
    $\forall i \in Dom(\tau^+(p))$,
    $\forall j\in [1,n]\setminus\{i\}$:
    $Var(s_i) \cap Var(s_j) =\emptyset$,
  \item \textrm{\bf(DN2)}
    $\forall \langle i \mapsto u_i\rangle \in \tau^+(p)$:
    $s_i$ is more general than $u_i$,
  \item \textrm{\bf(DN3)}
    $\forall i \in Dom(\tau^+(p))$,
    $\forall q(t_1,\dots,t_m) \in \B$,
    $\forall j\not\in Dom(\tau^+(q))$:
    $Var(s_i) \cap Var(t_j) =\emptyset,$
  \item \textrm{\bf(DN4)}
    $\forall q(t_1,\dots,t_m) \in \B$,
    $\forall \langle j \mapsto u_j\rangle \in \tau^+(q)$:
    $t_j$ is an instance of $u_j$.
  \end{itemize}
  A set of positions with associated terms is
  \emph{DN for a  program $P$} if it is DN for each clause of $P$.
\end{definition}

This definition says that any $s_i$ where $i$ is in
the domain of $\tau^+(p)$ (\ie{} position $i$ is
distinguished by $\tau^+$):
\textbf{(DN1)} does not
share its variables with the other arguments of
the head, \textbf{(DN2)} is more general than the term
$u_i$ that $i$ is mapped to by $\tau^+(p)$,
\textbf{(DN3)} distributes its variables to some arguments
$t_j$ of some atoms $q(t_1,\dots,t_m)$ in $\B$ such that
$j$ is in the domain of $\tau^+(q)$
(\ie{} position $j$ is distinguished by $\tau^+$).
Moreover, \textbf{(DN4)} says that any argument $t_j$,
where $j$ is distinguished by $\tau^+$,
of any atom $q(t_1,\dots,t_m)$ in $\B$ is such that
$t_j$ is an instance of the term $u_j$ that $j$ is
mapped to by $\tau^+(q)$.

\begin{example}
  [Example~\ref{example-set-of-pos-with-associated} continued]
  $\tau^+$ is DN for the following program:
  \begin{verbatim}
    p(f(x),y) :- q(x,g(x)), p(x,g(y))
    q(a,g(x)) :- q(a,g(b))
  \end{verbatim}
\end{example}

The preceding notion is closed under renaming:
\begin{proposition}\label{prop-DN}
  Let $c$ be a clause and $\tau^+$ be a set of positions
  with associated terms that is DN for $c$.
  Then $\tau^+$ is DN for every variant of $c$.
\end{proposition}
\begin{proof}
  The proof of this proposition is given in Appendix 1.
  \qed
\end{proof}

Notice that a set of positions is a particular set
of positions with associated terms in the following sense.
\begin{proposition}\label{prop-special-case}
  Let $\tau$ be a set of positions and $x$ be any variable.
  Let $\tau^+$ be the set of positions with associated
  terms defined as: for each $p\in\Pi$,
  $\tau^+(p) := (\; \tau(p) \rightarrowtail \{x\}\;)$.
  Then, the following holds.
  \begin{enumerate}
  \item \label{prop-special-case-item-one}
    An atom $A$ is $\Delta[\tau]$-more general than an
    atom $B$ iff $A$ is $\Delta[\tau^+]$-more general than
    $B$.
  \item \label{prop-special-case-item-two}
    For any clause $c$, $\tau$ is DN for $c$
    iff $\tau^+$ is DN for $c$.
  \end{enumerate}
\end{proposition}
\begin{proof}
  The proof follows from these remarks.
  \begin{itemize} 
  \item Item~\ref{prop-special-case-item-one}
    is a direct consequence of the definition
    of ``$\Delta$-more general'' (see
    Definition~\ref{definition-Delta-subsumes}) and
    the definition of the filter associated to a set of
    positions (see
    Definition~\ref{definition-set-of-pos-associated-filter})
    and to a set of positions with associated terms (see
    Definition~\ref{definition-associated-filter}).
  \item Item~\ref{prop-special-case-item-two}
    is a direct consequence of the definition
    of DN sets of positions (see
    Definition~\ref{definition-DN-set-of-positions}),
    and DN sets of positions with associated terms
    (see Definition~\ref{def-DN}). \qed
  \end{itemize}
\end{proof}

Finally, the sets of positions with associated terms
of Definition~\ref{def-DN} were named after
``DN'' because of the following result.
\begin{theorem}\label{theorem-DN-associated}
  Let $P$ be a logic program and 
  $\tau^+$ be a set of positions with associated terms that
  is DN for $P$. Then $\Delta[\tau^+]$ is DN for $P$.
\end{theorem}
\begin{proof}
    The proof of this theorem is given in Appendix 2. \qed
\end{proof}

For each predicate symbol $p$, let $\phi_p$ be
the function from $[1,\mathit{arity}(p)]$ to
$TU_{\mathcal{L}}$ whose domain is empty.
As in the case of sets of positions,
the set of DN sets of positions with associated terms
of a logic program is not empty as stated by
the following proposition.
\begin{proposition}
  Let $P$ be a logic program. Then
  $\tau^+_0 := \langle p \mapsto \phi_p \; | \; 
   p \in \Pi\rangle$
  is DN for $P$.
\end{proposition}
\begin{proof}
  By Definition~\ref{def-DN}.
  \qed
\end{proof}
Moreover, an atom $A$ is $\Delta[\tau^+_0]$-more
general than an atom $B$ iff $A$ is more general
than $B$.
So, in the context of the filter $\Delta[\tau^+_0]$,
the $\Delta$-Lifting Theorem~\ref{lifting-theorem}
is the same as the
Lifting Theorem~\ref{theorem-classical-lifting}.

\section{Examples}
\label{sect-examples}
In this section, we focus on \emph{left derivations} \ie{}
we only consider the leftmost selection rule:
$Q\lral Q'$ is a \emph{left derivation step} if it is
an SLD-derivation step whose selected atom is
the first atom of $Q$ from the left. We say that
\emph{a query $Q$ left loops w.r.t. a program $P$} if there
exists an infinite left derivation of $P\cup\{Q\}$.
Notice that the $\Delta$-Lifting Theorem~\ref{lifting-theorem}
provides a sufficient condition to identify left loops as
it generates the following corollary.
\begin{corollary}[from Theorem~\ref{lifting-theorem}]
  \label{theorem-loop-detection}
  Let $P$ be a logic program and $\Delta$ be a DN filter
  for $P$. If $A \lral^*_P B_1,\B_1$ and
  $B_1 \lral^+_P B_2,\B_2$ and
  $B_2$ is $\Delta$-more general than $B_1$
  then $P \cup \{A\}$ left loops.
\end{corollary}
\begin{proof}
  The proof is similar to that of
  Corollary~\ref{corollary-loop-detection}.
\end{proof}

This section presents some examples of atomic
queries that have an infinite left derivation w.r.t.
a given logic program.
We use Corollary~\ref{theorem-loop-detection}
with DN sets of positions and DN sets of positions
with associated terms. In each case, it is not possible
to conclude using the classical Lifting
Theorem~\ref{theorem-classical-lifting}. The last
example exhibits a case where we are not able conclude.

\begin{example}
  Let $\Pi := \{p\}$ with $p$ a relation symbol whose
  arity equals 2 and let 
  \[c := p(f(x),y) \leftarrow p(x,g(y))\;.\]
  Then
  \[\tau := \langle p \mapsto \{2\} \rangle\] is
  a DN set of positions for $c$.
  The filter associated to $\tau$
  (see Definition~\ref{definition-set-of-pos-associated-filter})
  is
  \[\Delta[\tau] := \langle p \mapsto \langle 2 \mapsto
    f_{\mathit{true}} \rangle \rangle\;.\]
  Notice that, by Theorem~\ref{theorem-set-of-positions},
  \[\Delta[\tau] \textrm{ is DN for } \{c\}\;.\]
  Moreover, from the head $p(f(x),y)$ of $c$ 
  we get
  \[p(f(x),y) \lral_c p(x',g(y'))\;.\]
  Applying Corollary~\ref{theorem-loop-detection},
  as  $p(x',g(y'))$ is
  $\Delta[\tau]$-more general than $p(f(x),y)$,
  we get that
  \[p(f(x),y) \textrm{ left loops w.r.t. } \{c\}\;.\]
  We point out that we do not get this
  result from the classical Lifting
  Theorem~\ref{theorem-classical-lifting}
  as $p(x',g(y'))$ is not more general than $p(f(x),y)$.

  By the $\Delta$-Lifting Theorem~\ref{lifting-theorem},
  we can also conclude that each query that is
  $\Delta[\tau]$-more general than $p(f(x),y)$ also left
  loops w.r.t. $\{c\}$. This means that
  each query of form $p(t_1,t_2)$ where $t_1$ is a term that
  is more general than $f(x)$ and $t_2$ is \emph{any term}
  (because $\tau(p) = \{2\}$)
  left loops w.r.t. $\{c\}$.
  \qed
\end{example}

\begin{example}
  Let $\Pi := \{p\}$ with $p$ a relation symbol whose
  arity equals 2 and let 
  \[c := p(f(x),g(y)) \leftarrow p(x,g(h(y))) \; .\]
  Notice that from the head $p(f(x),g(y))$ of $c$
  we get
  \[p(f(x),g(y)) \lral_c p(x',g(h(y'))) \; .\]

  The only DN set of positions for
  $c$ is $\tau_0 := \langle p \mapsto \emptyset \rangle$
  because each argument of the head of $c$ is not a
  variable (see
  Definition~\ref{definition-DN-set-of-positions}).
  Hence, as $p(x',g(h(y')))$ is not $\Delta[\tau_0]$-more
  general than $p(f(x),g(y))$, we can not conclude
  using $\tau_0$ that 
  $p(f(x),g(y))$ left loops w.r.t. $\{c\}$.

  However,
  \[\tau^+ := \langle \; p \mapsto \langle 2 \mapsto g(y)
    \rangle \; \rangle\]
  is a set of positions with associated terms that is DN
  for $\{c\}$. Hence, by Theorem~\ref{theorem-DN-associated},
  the associated filter $\Delta[\tau^+]$
  (see Definition~\ref{definition-associated-filter})
  is DN for $\{c\}$. As $p(x',g(h(y')))$ is
  $\Delta[\tau^+]$-more general than $p(f(x),g(y))$,
  by Corollary~\ref{theorem-loop-detection}
  we get that
  \[p(f(x),g(y)) \textrm{ left loops w.r.t. } \{c\}\;.\]
  By the $\Delta$-Lifting Theorem~\ref{lifting-theorem}, we can
  also conclude that each query that is $\Delta[\tau^+]$-more
  general than
  $p(f(x),g(y))$ also left loops w.r.t. $\{c\}$. This means that
  each query of form $p(t_1,t_2)$ where $t_1$ is a term that is
  more general than $f(x)$ and $t_2$ is
  \emph{any instance of $g(y)$} (because
  $\tau^+(p) = \langle 2\mapsto g(y) \rangle$)
  left loops w.r.t. $\{c\}$.
  \qed
\end{example}

\begin{example}
  Let $\Pi := \{p,q\}$ with $p$ and $q$ two relation symbols
  whose arity equals 2. 
  Let 
  \[\begin{array}{r@{\;:=\;}r@{\;\leftarrow\;}l}
      c_1 & p(f(x),y) & q(x,g(x)),\; p(x,g(y)) \\[1ex]
      c_2 & q(a,g(x)) & q(a,g(b))
  \end{array}\]
  Then
  \[\tau^+ := \langle \; p \mapsto \langle 2 \mapsto y
    \rangle,\;  \; q \mapsto \langle 2 \mapsto g(x) \rangle
    \; \rangle\]
  is a set of positions with associated terms that is DN
  for $\{c_1,c_2\}$. Hence, by
  Theorem~\ref{theorem-DN-associated},
  the associated filter $\Delta[\tau^+]$
  (see Definition~\ref{definition-associated-filter})
  is DN for $\{c_1,c_2\}$. Moreover, from the head
  $q(a,g(x))$ of $c_2$ we get
  \[q(a,g(x)) \lral_{c_2} q(a,g(b))\;.\]
  Applying Corollary~\ref{theorem-loop-detection}, as
  $q(a,g(b))$ is $\Delta[\tau^+]$-more general than
  $q(a,g(x))$, we get that
  \[q(a,g(x)) \textrm{ left loops w.r.t. } c_2\;.\]
  By the $\Delta$-Lifting Theorem~\ref{lifting-theorem},
  each query of form $q(t_1,t_2)$ where $t_1$ is a term that
  is more general than $a$ and $t_2$ is
  \emph{any instance of $g(x)$} (because
  $\tau^+(q) = \langle 2\mapsto g(x) \rangle$)
  left loops w.r.t. $c_2$.
  Notice that from the head $p(f(x),y)$ of $c_1$ 
  we get
  \[p(f(x),y) \lral_{c_1} q(x',g(x')), p(x',g(y'))\;.\]
  As $q(x',g(x'))$ is such that $x'$ is more general
  than $a$ and $g(x')$ is an instance of $g(x)$,
  we get that $q(x',g(x'))$ left loops w.r.t. $c_2$.
  Consequently, $p(f(x),y)$ left loops w.r.t.
  $\{c_1,c_2\}$.
  So, again by the $\Delta$-Lifting
  Theorem~\ref{lifting-theorem},
  each query of form $p(t_1,t_2)$, where $t_1$ is a term
  that is more general than $f(x)$ and $t_2$ is
  \emph{any instance of $y$} (because
  $\tau^+(p) = \langle 2\mapsto y \rangle$)
  left loops w.r.t. $\{c_1,c_2\}$.
  \qed
\end{example}

\begin{example}\label{example-does-not-work}
  Let $\Pi := \{p\}$ with $p$ a relation symbol whose
  arity equals 2 and let 
  \[c := p(x,x) \leftarrow p(f(x),f(x)) \; .\]
  As the arguments
  of the head of $c$ have one common variable $x$, the
  only set of positions with associated terms that is
  DN for $c$ is $\tau^+_0$ such that the domain of
  $\tau^+_0(p)$ is empty (see \textbf{(DN1)}
  in Definition~\ref{def-DN}).
  Notice that from the head $p(x,x)$ of $c$ we get
  \[\begin{array}{l}
    p(x,x) \lral_c p(f(x_1),f(x_1)) \cdots\\[1ex]
    \hspace{1.5cm}\cdots p(f^n(x_n),f^n(x_n)) \lral_c%
    p(f^{n+1}(x_{n+1}),f^{n+1}(x_{n+1}))%
    \cdots
  \end{array}\]
  As the arguments of $p$ grow from
  step to step, there cannot be any query in the derivation
  that is $\Delta[\tau^+_0]$-more general than
  one of its ancestors.
  Consequently, we can not conclude that $p(x,x)$ left loops
  w.r.t $c$.
  \qed
\end{example}

\section{Related Works}
\label{section-related-works}
Some extensions of the Lifting Theorem
with respect to infinite derivations are presented in
\cite{Gori97a}, where  the authors 
study numerous properties of finite failure. 
The non-ground finite failure set 
of a logic program  is defined as the set of possibly non-ground atoms
which admit a fair finitely failed SLD-tree w.r.t. the program. 
This denotation is shown correct in the following sense. 
If two programs have the same non-ground finite failure set, then  any
ground or non-ground 
goal which finitely fails w.r.t.  one program also finitely fails w.r.t. the other.
Such a property is false when we consider the standard ground
finite failure set. 
The proof
of correctness of the non-ground finite failure semantics relies on
the following result. First, a derivation is 
called non-perpetual if it is a fair infinite 
derivation and there  exists  a finite depth from which unfolding
does not instantiate the original goal any more.
Then the authors define the definite answer goal of
a non-perpetual derivation as the maximal instantiation
of the original goal. A crucial lemma states that any instance of 
the definite answer goal admits a similar non-perpetual derivation.
Compared to our work, note that we do need fairness as an hypothesis
for  the $\Delta$-Lifting Theorem~\ref{lifting-theorem}. 
On the other hand, investigating the relationships between 
non-ground arguments of definite answer and 
neutral arguments is an interesting problem.
\\
\\
Loop checking in logic programming 
is also a subject related to our work. In this area,
\cite{Bol91b} sets up some solid foundations.
A loop check is a device to prune
derivations when it seems appropriate. A loop checker
is defined as \emph{sound} if no solution is lost.
It is \emph{complete} if all infinite derivations
are pruned. A complete loop check may
also prune finite derivations. 
The authors shows that even for function-free
programs (also known as Datalog programs), 
sound and complete loop checks are out of reach.
Completeness is shown 
only for some restricted classes of function-free programs. 

We now review loop checking in more details.
To our best knowledge, among all existing loop
checking mechanisms only OS-check \cite{Sahlin90},
EVA-check \cite{Shen97} and VAF-check \cite{Shen01a}
are suitable for logic programs with function symbols.
They rely on a structural characteristic
of infinite SLD-derivations, namely, the
growth of the size of some generated subgoals.
This is what the following theorem states.
\begin{theorem}
  Consider an infinite SLD-derivation $\xi$
  where the leftmost selection
  rule is used. Then there are infinitely many queries
  $Q_{i_1}$, $Q_{i_2}$, \dots{}
  (with $i_1<i_2<\dots$) 
  in $\xi$
  such that for any $j\geq 1$,
  the selected atom $A_{i_j}$ of $Q_{i_j}$ is an ancestor
  of the selected atom $A_{i_{j+1}}$ of $Q_{i_{j+1}}$ and
  $\mathit{size}(A_{i_{j+1}}) \geq \mathit{size}(A_{i_j})$.
\end{theorem}
Here, $\mathit{size}$ is a given function that maps an atom
to its size which is defined in terms of the number of
symbols appearing in the atom.
As this theorem does not provide any sufficient condition
to detect infinite SLD-derivations,
the three loop checking mechanisms mentioned above may
detect finite derivations as infinite.
However, these mechanisms are
\emph{complete w.r.t. the leftmost selection rule} \ie{}
they detect all infinite loops when the leftmost selection
rule is used.

OS-check (for OverSize loop check)
was first introduced by Shalin \cite{Sahlin90,Sahlin93a}
and was then formalized by Bol \cite{Bol93}.
It is based on a function $\mathit{size}$ that can
have one of the three following definitions: for any atoms
$A$ and $B$, either  $\mathit{size}(A)=\mathit{size}(B)$,
either $\mathit{size}(A)$ (resp. $\mathit{size}(B)$) is
the count of symbols appearing in $A$ (resp. $B$), either  
$\mathit{size}(A)\leq\mathit{size}(B)$ if for each $i$,
the count of symbols of the $i$-th argument of $A$
is smaller than or equal to that of the $i$-th argument of
$B$.
OS-check says that an SLD-derivation may be infinite
if it generates an atomic subgoal $A$ that is \emph{oversized},
\ie{} that has ancestor subgoals which have
the same predicate symbol as $A$ and whose size is smaller
than or equal to that of $A$.

EVA-check (for Extented Variant Atoms loop check) was
introduced by Shen \cite{Shen97}. It is based on the notion
of \emph{generalized variants}. EVA-check says that an
SLD-derivation may be infinite if it generates an atomic
subgoal $A$ that is a generalized variant of some of its
ancestor $A'$, \ie{} $A$ is a variant of $A'$ except for
some arguments whose size increases from $A'$ to $A$ via
a set of recursive clauses. Here the size function that is
used applies to predicate arguments, \ie{} to terms,
and it is fixed: it is defined as the the count of
symbols that appear in the terms.
EVA-check is more reliable
than OS-check because it is less likely to
mis-identify infinite loops \cite{Shen97}.
This is mainly due to the fact that, unlike OS-check,
EVA-check refers to the informative internal structure
of subgoals.

VAF-check (for Variant Atoms loop check for logic programs
with Functions) was proposed by Shen \emph{et al.}
\cite{Shen01a}. It is based on the notion
of \emph{expanded variants}.
VAF-check says that an SLD-derivation
may be infinite if it generates an atomic subgoal $A$ that is
an expanded variant of some of its ancestor $A'$, \ie{}
$A$ is a variant of $A'$ except for some arguments 
$t_{i_1}, \dots, t_{i_n}$ such that:
$t_{i_1}$ grows from $A'$ to $A$ into a function containing
$t_{i_1}$, \dots, $t_{i_n}$ grows from $A'$ to $A$ into a
function containing $t_{i_n}$. VAF-check is as reliable
as and more efficient than EVA-check \cite{Shen01a}.

The main difference with our work is that
we want to pinpoint \emph{some} infinite derivations,
based on syntactical properties of the program.
We are not interested in completeness nor in soundness. 
Notice, however, that using the $\Delta$-Lifting
Theorem~\ref{lifting-theorem} as a loop checker leads to
a device that is neither complete (see
Example~\ref{example-does-not-work}) nor sound
since the Lifting Theorem~\ref{theorem-classical-lifting}
is a particular case of the $\Delta$-Lifting
Theorem~\ref{lifting-theorem}.


\section{Conclusion}

We have presented a generalization of the lifting lemma
for logic programming, which allows to disregard 
some arguments, termed neutral arguments, 
while checking for subsumption. We have
investigated the model theoretic consequence of this
generalization and have proposed two syntactic criteria
for statically identifying neutral arguments.

A first  application of this work has already been
presented in \cite{Mesnard02a}, in the area of termination analysis
of logic programs. We combine cTI \cite{Mesnard01}, a termination inference
tool with a non-termination inference analyzer whose correctness
relies on our generalized lifting lemma. 
The resulting combined
analysis may sometimes \emph{characterize} the termination behavior
of some concrete logic program w.r.t. to the left selection rule and
the language we use to describe classes of queries.

Finally, this paper leaves  numerous questions open. For instance,
it might be interesting to try to generalize this approach to constraint
logic programming \cite{JL87}. Can we obtain higher
level proofs compared to those we give? Can we propose more
abstract or semantically-based criteria for identifying neutral arguments?
\\
\\

{\bf Acknowledgements.} We thank Ulrich Neumerkel for
initial discussions on this topic and Roberto Bagnara
for interesting suggestions.

\bibliography{PayetMesnard-TOCL}

\begin{thebibliography}{10}

\bibitem{Apt97a}
K.~R. Apt.
\newblock {\em From {Logic Programming} to {P}rolog}.
\newblock Prentice Hall, 1997.

\bibitem{Apt82a}
K.~R. Apt and M.~H. {Van Emden}.
\newblock Contributions to the theory of logic programming.
\newblock {\em Journal of the ACM}, 29(3):841--862, 1982.

\bibitem{Bol93}
R.~N. Bol.
\newblock Loop checking in partial deduction.
\newblock {\em Journal of {L}ogic {P}rogramming}, 16:25--46, 1993.

\bibitem{Bol91b}
R.~N. Bol, K.~R. Apt, and J.~W. Klop.
\newblock An analysis of loop checking mechanisms for logic programs.
\newblock {\em Theoretical {C}omputer {S}cience}, 86:35--79, 1991.

\bibitem{Chang73}
C.-L. Chang and R.~Lee.
\newblock {\em Symbolic Logic and Mechanical Theorem Proving}.
\newblock Computer Science Classics. Academic Press, 1973.

\bibitem{clark79a}
K.~L. Clark.
\newblock Predicate logic as a computational formalism.
\newblock Technical Report Doc 79/59, {Logic Programming} Group, Imperial
  College, London, 1979.

\bibitem{Deransart87}
P.~Deransart and G.~Ferrand.
\newblock Programmation en logique avec n\'egation: pr\'esentation formelle.
\newblock Technical Report 87/3, Laboratoire d'Informatique, D\'epartement de
  Math\'ematiques et d'Informatique, Universit\'e d'Orl\'eans, 1987.

\bibitem{Doe93}
H.~C. Doets.
\newblock Levationis laus.
\newblock {\em Journal of {L}ogic and {C}omputation}, 3(5):487--516, 1993.

\bibitem{Gori97a}
R.~Gori and G.~Levi.
\newblock Finite failure is and-compositional.
\newblock {\em Journal of Logic and Computation}, 7(6):753--776, 1997.

\bibitem{JL87}
J.~Jaffar and J.~L. Lassez.
\newblock Constraint logic programming.
\newblock In {\em Proc. of the ACM Symposium on Principles of Programming
  Languages}, pages 111--119. {ACM} {P}ress, 1987.

\bibitem{Lloyd87a}
J.~W. Lloyd.
\newblock {\em Foundations of {Logic Programming}}.
\newblock Springer-Verlag, 1987.

\bibitem{Falaschi93}
M.~Martelli M.~Falaschi, G.~Levi and C.~Palamidessi.
\newblock A model-theoretic reconstruction of the operational semantics of
  logic programs.
\newblock {\em Information and Computation}, 102(1):86--113, 1993.

\bibitem{Mesnard01}
F.~Mesnard and U.~Neumerkel.
\newblock Applying static analysis techniques for inferring termination
  conditions of logic programs.
\newblock In P.~Cousot, editor, {\em Static Analysis Symposium}, volume 2126 of
  {\em {LNCS}}, pages 93--110, Berlin, 2001. Springer-{V}erlag.

\bibitem{Mesnard02a}
F.~Mesnard, E.~Payet, and U.~Neumerkel.
\newblock Detecting optimal termination conditions of logic programs.
\newblock In M.~Hermenegildo and G.~Puebla, editors, {\em Proc. of the 9th
  International Symposium on Static Analysis}, volume 2477 of {\em Lecture
  {N}otes in {C}omputer {S}cience}, pages 509--525. Springer-{V}erlag,
  {B}erlin, 2002.

\bibitem{Okeefe90a}
R.~O'Keefe.
\newblock {\em The Craft Of Prolog}.
\newblock MIT Press, 1990.

\bibitem{Rob65}
J.~A. Robinson.
\newblock A machine-oriented logic based on the resolution principle.
\newblock {\em Journal of the ACM}, 12(1):23--41, 1965.

\bibitem{Sahlin90}
D.~Sahlin.
\newblock The mixtus approach to automatic partial evaluation of full prolog.
\newblock In S.~Debray and M.~Hermenegildo, editors, {\em Proc. of the 1990
  North American Conference on Logic Programming}, pages 377--398. MIT Press,
  Cambridge, MA, 1990.

\bibitem{Sahlin93a}
D.~Sahlin.
\newblock Mixtus: an automatic partial evaluator for full {P}rolog.
\newblock {\em New {G}eneration {C}omputing}, 12(1):7--51, 1993.

\bibitem{Shen97}
Y-D. Shen.
\newblock An extended variant of atoms loop check for positive logic programs.
\newblock {\em New {G}eneration {C}omputing}, 15(2):187--204, 1997.

\bibitem{Shen01a}
Y-D. Shen, L-Y. Yuan, and J-H. You.
\newblock Loops checks for logic programs with functions.
\newblock {\em Theoretical {C}omputer {S}cience}, 266(1-2):441--461, 2001.

\end{thebibliography}

\section{Appendix 1}

\subsection{Proof of Proposition~\ref{prop-var-included}}
\begin{description}
\item[$\Leftarrow$] By definition.
\item[$\Rightarrow$]
  As $Q$ is $\Delta$-more general than $Q'$,
  there exists a substitution
  $\sigma$ such that $Q$ is $\Delta$-more general than $Q'$
  for $\sigma$.
  Notice that $Q$ is also $\Delta$-more general than $Q'$ for
  the substitution obtained by restricting the domain of
  $\sigma$ to the variables appearing the positions of
  $Q$ not distinguished by $\Delta$.
  More precisely, let
  \[T := \{t \in TU_{\mathcal{L}} \; | \;
  \exists p(t_1,\dots,t_n) \in Q, \;
  \exists i \in [1,n]\setminus Dom(\Delta(p)), \; t=t_i\}\]
  and
  \[\eta := \sigma | Var(T) \; .\]
  Then, $Dom(\eta) \subseteq Var(T)$ \ie{}
  \begin{equation}\label{prop-var-included-eq1}
    Dom(\eta) \subseteq Var(Q)
  \end{equation}
  and $Q$ is $\Delta$-more general than $Q'$ for $\eta$.
  
  Now, let $x \in Dom(\eta)$. Then, as 
  $Dom(\eta) \subseteq Var(T)$,
  there exists $A := p(t_1,\dots,t_n) \in Q$
  and $i \in [1,n]\setminus Dom(\Delta(p))$ such that
  $x \in Var(t_i)$.
  
  As $Q$ is $\Delta$-more general than $Q'$ for $\eta$,
  there exists
  $A':=p(t'_1,\dots,t'_n)$ in $Q'$ such that $A$ is
  $\Delta$-more general than $A'$ for $\eta$.
  But, as $i \in [1,n]\setminus Dom(\Delta(p))$, we have
  $t'_i=t_i\eta$.
  So, as $x \in Var(t_i)$, $x\eta$ is a subterm of $t'_i$.
  Consequently, $Var(x\eta) \subseteq Var(t'_i)$, so
  $Var(x\eta) \subseteq Var(Q')$.
  
  So, we have proved that for each $x\in Dom(\eta)$,
  $Var(x\eta) \subseteq Var(Q')$, \ie{} we have proved that
  \begin{equation}\label{prop-var-included-eq2}
    Ran(\eta)\subseteq Var(Q') \;.
  \end{equation}
  Finally, (\ref{prop-var-included-eq1}) and
  (\ref{prop-var-included-eq2})
  imply that $Dom(\eta) \cup Ran(\eta) \subseteq Var(Q,Q')$
  \ie{} that
  \[Var(\eta) \subseteq Var(Q,Q')\; .\]
  \qed
\end{description}

\subsection{Proof of Proposition~\ref{prop-DN}}
Let $c := p(s_1,\dots,s_n) \leftarrow \B$ and
$c' := p(s'_1,\dots,s'_n) \leftarrow \B'$ be a variant of $c$.
Then, there exists a renaming
$\gamma$ such that $c'=c\gamma$.
\begin{description}
\item[\textbf{(DN1)}] Let $i\in Dom(\tau^+(p))$.
  Suppose that there exists $j \neq i$ such that
  $Var(s'_i) \cap Var(s'_j)\neq\emptyset$ and let us derive a contradiction.
  
  Let $x'\in Var(s'_i) \cap Var(s'_j)$. As $s'_j=s_j\gamma$, there exists
  $x \in Var(s_j)$ such that $x'=x\gamma$.
  
  For such an $x$,
  as $j \neq i$ and as $Var(s_i) \cap Var(s_j)=\emptyset$
  (because $\tau^+$ is DN for $c$), we have $x \not\in Var(s_i)$.
  So, as $\gamma$ is a 1-1 and onto
  mapping from its domain to itself, we have
  $x\gamma \not\in Var(s_i\gamma)$%
  \footnote{\label{footnote-proof-prop-DN}
    Because if $x\gamma\in Var(s_i\gamma)$, then either
    $x\in Var(s_i)$,
    either $x\gamma \in Var(s_i)$ and $(x\gamma)\gamma = x\gamma$. The
    former case is impossible because we said that $x\not\in Var(s_i)$.
    The latter case is impossible too because $(x\gamma)\gamma= x\gamma$
    implies that $x\gamma \not\in Dom(\gamma)$ \ie{}
    $x\not\in Dom(\gamma)$ (because $\gamma$ is a 1-1 and onto
    mapping from its domain to itself); so,
    $x = x\gamma$ \ie{}, as $x\gamma\in Var(s_i)$,
    $x\in Var(s_i)$.
  }, \ie{} $x' \not\in Var(s'_i)$.
  Contradiction!
  
  Consequently, $Var(s'_i) \cap Var(s'_j)=\emptyset$.
\item[\textbf{(DN2)}] Let
  $\langle i \mapsto u_i\rangle\in\tau^+(p)$.
  As $s_i$ is more general than $u_i$
  (because $\tau^+$ is DN for $c$) and as $s'_i$ is a variant of
  $s_i$, $s'_i$ is more general than $u_i$.
\item[\textbf{(DN3)}] Let $i\in Dom(\tau^+(p))$.
  Suppose that there exists
  $q(t'_1,\dots,t'_m) \in \B'$ and
  $j \not\in Dom(\tau^+(q))$
  such that $Var(s'_i) \cap Var(t'_j)\neq \emptyset$. Let us
  derive a contradiction.
  
  Let $x'\in Var(s'_i) \cap Var(t'_j)$.
  As $\B'=\B\gamma$, there exists
  $q(t_1,\dots,t_m) \in \B$ such that
  $q(t'_1,\dots,t'_m)=q(t_1,\dots,t_m)\gamma$, \ie{}
  $t'_j=t_j\gamma$. So, as $x' \in Var(t'_j)$, there exists
  $x \in Var(t_j)$ such that  $x'=x\gamma$.
  
  For such an $x$, as the elements of $Var(s_i)$ only occur
  in those $t_k$ such that
  $k \in Dom(\tau^+(q))$ (because $\tau^+$ is DN for $c$) and
  as $x \in Var(t_j)$ with
  $j \not\in Dom(\tau^+(q))$, we have $x \not\in Var(s_i)$.
  So, as $\gamma$ is a 1-1 and onto
  mapping from its domain to itself, we have
  $x\gamma \not\in Var(s_i\gamma)$
  (see footnote~\ref{footnote-proof-prop-DN}),
  \ie{} $x' \neq Var(s'_i)$.
  Contradiction!
  
  Therefore, for each $q(t'_1,\dots,t'_m) \in \B'$ and
  $j \not\in Dom(\tau^+(q))$, we have
  $Var(s'_i) \cap Var(t'_j) = \emptyset$.
\item[\textbf{(DN4)}] Let $q(t'_1,\dots,t'_m) \in \B'$ and
  $\langle j \mapsto u_j\rangle \in \tau^+(q)$.
  As $t_j$ is an instance of
  $u_j$ (because $\tau^+$ is DN for $c$) and as $t'_j$ is a
  variant of $t_j$, $t'_j$ is an instance of $u_j$.
\end{description}
Finally, we have established that $\tau^+$ is DN for $c'$.
\qed

\section{Appendix 2: DN Sets of Positions with Associated
  Terms Generate DN Filters}
\label{section-technical-proofs}

In this section, we give a proof of
Theorem~\ref{theorem-DN-associated}.

\subsection{Context}
%
All the results of this section are parametric to the
following context:
\begin{itemize}
\item $\tau^+$ denotes a set of positions with associated
  terms that is DN for a program $P$,
\item $Q \lral^{\theta}_c Q_1$ is an SLD-derivation step
  where
  \begin{itemize}
  \item $c \in P$,
  \item $Q:=\A,p(t_1,\dots,t_n),\C$ where
    $p(t_1,\dots,t_n)$ is the selected atom,
  \item $c_1:=p(s_1,\dots,s_n)\leftarrow \B$ is the input
    clause used,
  \end{itemize}
\item $Q':=\A',p(t'_1,\dots,t'_n),\C'$ is
  $\Delta[\tau^+]$-more general than $Q$ \ie{}, by
  Proposition~\ref{prop-var-included}, there exists a
  substitution $\eta$ such that
  $Var(\eta) \subseteq Var(Q,Q')$ and $Q'$ is
  $\Delta[\tau^+]$-more general than
  $Q$ for $\eta$. Moreover, the position of
  $p(t'_1,\dots,t'_n)$ in $Q'$ is the same as that of
  $p(t_1,\dots,t_n)$ in $Q$.
\end{itemize}

\subsection{Technical Definitions and Lemmas}

\begin{definition}[Technical Definition]\label{definition-technical}
  Let $c'_1:=p(s'_1,\dots,s'_n)\leftarrow \B'$ be a clause such that
  \begin{itemize}
  \item $Var(c'_1) \cap Var(Q,Q')=\emptyset$ and
  \item $c_1=c'_1\gamma$ for some renaming
    $\gamma$ satisfying $Var(\gamma)\subseteq Var(c_1,c'_1)$.
  \end{itemize}
  As $c'_1$ is a variant of $c_1$ and $c_1$ is a variant
  of $c$, then $c'_1$ is a variant of $c$. Moreover,
  as $\tau^+$ is DN for $c$, by Proposition~\ref{prop-DN},
  $\tau^+$ is DN for $c'_1$.
  So, by \textbf{(DN2)} in Definition~\ref{def-DN},
  for each $\langle i\mapsto u_i\rangle \in \tau^+(p)$
  there exists a substitution $\delta_i$ such that
  $u_i=s'_i\delta_i$.

  Moreover, as $p(t'_1,\dots,t'_n)$ is $\Delta[\tau^+]$-more
  general than $p(t_1,\dots,t_n)$, for each
  $\langle i\mapsto u_i\rangle \in \tau^+(p)$,
  $t'_i$ is an instance of $u_i$.
  So, there exists a substitution $\delta'_i$
  such that $t'_i=u_i\delta'_i$.

  For each $i \in Dom(\tau^+(p))$, we set
  \[\sigma_i \stackrel{def}{=} (\delta_i\delta'_i)|Var(s'_i)\; .\]
  Moreover, we set:
  \[\sigma \stackrel{def}{=} \bigcup\limits_{i \in Dom(\tau^+(p))} 
    \sigma_i\;.\]
\end{definition}

\begin{lemma}
  The set $\sigma$ of Definition~\ref{definition-technical}
  is a well-defined substitution.
\end{lemma}
\begin{proof}
  Notice that, as $\tau^+$ is DN for $c'_1$,
  by \textbf{(DN1)} in Definition~\ref{def-DN} we have
  \[
    \forall i \in Dom(\tau^+(p)), \;
    \forall j \in [1,n]\setminus\{i\}, \;
    Var(s'_i)\cap Var(s'_j)=\emptyset\;.
  \]
  Consequently, 
  \[\forall i,j \in Dom(\tau^+(p)), \;
    i \neq j \Rightarrow
    Dom(\sigma_i) \cap Dom(\sigma_j)=\emptyset\;.\]
  Moreover, for each $i \in Dom(\tau^+(p))$,
  $\sigma_i$ is a well-defined substitution.
  So, $\sigma$ is a well-defined substitution.
  \qed
\end{proof}

\begin{lemma}[Technical Lemma]\label{technical-lemma}
  Let $c'_1:=p(s'_1,\dots,s'_n)\leftarrow \B'$ be a clause
  such that
  \begin{itemize}
  \item $Var(c'_1) \cap Var(Q,Q')=\emptyset$ and
  \item $c_1=c'_1\gamma$ for some renaming
    $\gamma$ satisfying $Var(\gamma)\subseteq Var(c_1,c'_1)$.
  \end{itemize}

  Let $\sigma$ be the substitution of
  Definition~\ref{definition-technical}.
  Then, the substitution $\sigma\eta\gamma\theta$ is a
  unifier of $p(t'_1,\dots,t'_n)$ and $p(s'_1,\dots,s'_n)$.
\end{lemma}
\begin{proof}
  The result follows from the following facts.
  \begin{itemize}
  \item For each $\langle i\mapsto u_i\rangle \in \tau^+(p)$,
    we have:
    \[s'_i\sigma =s'_i\sigma_i=s'_i\delta_i\delta'_i=
      (s'_i\delta_i)\delta'_i=u_i\delta'_i=t'_i\]
    and $t'_i\sigma=t'_i$ because
    $Dom(\sigma)\subseteq Var(c'_1)$ and
    $Var(Q')\cap Var(c'_1)=\emptyset$.
    So, $s'_i\sigma=t'_i\sigma$ and
    $s'_i\sigma\eta\gamma\theta=t'_i\sigma\eta\gamma\theta$.
  \item For each $i \in [1,n]\setminus Dom(\tau^+(p))$, we have:
    \[s'_i\eta\gamma\theta=(s'_i\eta)\gamma\theta=
    s'_i\gamma\theta=(s'_i\gamma)\theta=s_i\theta\]
    and
    \[t'_i\eta\gamma\theta=(t'_i\eta)\gamma\theta=
      t_i\gamma\theta=(t_i\gamma)\theta=t_i\theta\]
    and $s_i\theta=t_i\theta$ because $\theta$ is a unifier of
    $p(s_1,\dots,s_n)$ and $p(t_1,\dots,t_n)$
    (because $Q\lral^{\theta}_c Q_1$ with $c_1$ as input clause used).
    So,
    \begin{equation}\label{tech-lemma-eq1}
      s'_i\eta\gamma\theta=t'_i\eta\gamma\theta
    \end{equation}
  \item For each $i \in [1,n]\setminus Dom(\tau^+(p))$, we also have:
    \begin{itemize}
    \item $s'_i\sigma=s'_i$ because
      $Dom(\sigma)=Var\big(\{s'_j \; | \; j \in Dom(\tau^+(p))\}\big)$
      and, by \textbf{(DN1)} in Definition~\ref{def-DN},
      $Var\big(\{s'_j \; | \; j \in Dom(\tau^+(p))\}\big)
      \cap Var(s'_i) = \emptyset$;
    \item $t'_i\sigma=t'_i$ because $Dom(\sigma)\subseteq Var(c'_1)$
      and $Var(Q')\cap Var(c'_1)=\emptyset$.
    \end{itemize}
    Therefore, because of~(\ref{tech-lemma-eq1}),
    $s'_i\sigma\eta\gamma\theta=t'_i\sigma\eta\gamma\theta$.
    \qed
  \end{itemize}
\end{proof}

\subsection{$\Delta$-Propagation}

Now we extend the following Propagation Lemma
that is proved by Apt in \cite{Apt97a} p.~54--56. 
\begin{lemma}[Propagation]
  \label{lemma-classical-propagation}
  Let $G$, $G_1$, $G'$ and $G'_1$ be queries such that
  $G\lral_c G_1$ and $G' \lral_c G'_1$ and:
  \begin{itemize}
  \item $G$ is an instance of $G'$
  \item in $G$ and $G'$ atoms in the same positions
    are selected.
  \end{itemize}
  Then, $G_1$ is an instance of $G'_1$.
\end{lemma}

First we establish the following result.
\begin{lemma}\label{lemma-propagation}
  Suppose that there exists an SLD-derivation step of form
  $Q' \lral^{\theta'}_c Q'_1$ where $p(t'_1,\dots,t'_n)$
  is the selected atom and the input clause is $c'_1$
  such that $Var(Q) \cap Var(c'_1) = \emptyset$.
  Then, $Q'_1$ is $\Delta[\tau^+]$-more general
  than $Q_1$.
\end{lemma}
\begin{proof}
  Notice that we have
  \[Var(Q) \cap Var (c_1) = Var(Q,Q') \cap Var(c'_1)
    = \emptyset \; .\]
  Moreover, as $c_1$ is a variant of $c'_1$,
  there exists a renaming $\gamma$ such that
  \[Var(\gamma) \subseteq Var(c_1,c'_1)
    \text{\quad and \quad} c_1=c'_1\gamma \; .\]
  Let $c'_1:=p(s'_1,\dots,s'_n)\leftarrow \B'$. Then,
  \[Q_1 = (\A,\B,\C)\theta \text{\quad and \quad}
  Q'_1 = (\A',\B',\C')\theta' \; .\]
  $\tau^+$ is DN for $c$ and $c'_1$ is a variant of $c$. So,
  by Proposition~\ref{prop-DN}, $\tau^+$ is DN for $c'_1$.
  Let $\sigma$ be the substitution of
  Definition~\ref{definition-technical}.
  \begin{itemize}
  \item Let $A':=q(v'_1,\dots,v'_m) \in \A'$. As $\A'$ is
    $\Delta[\tau^+]$-more general than $\A$ for $\eta$,
    $\A'$ and $\A$ have the same length.
    Moreover, if $k$ denotes the position of $A'$ in $\A'$,
    then the $k^{\text{th}}$ atom of $\A$ has form
    $q(v_1,\dots,v_m)$.
    \begin{itemize}
    \item As $A'$ is $\Delta[\tau^+]$-more general than $A$
      for $\eta$, for each
      $\langle j\mapsto u_j\rangle \in \tau^+(q)$,
      $v'_j$ is an instance of $u_j$.
    \item For each $j \in [1,m]\setminus Dom(\tau^+(q))$ we have:
      \[v'_j\sigma\eta\gamma\theta=(v'_j\sigma)\eta\gamma\theta
      =v'_j\eta\gamma\theta\]
      because
      $Dom(\sigma)=Var\big(\{s'_i \; | \; i \in
       Dom(\tau^+(p))\}\big)\subseteq Var(c'_1)$
      and $Var(c'_1)\cap Var(Q')=\emptyset$. Moreover,
      \[v'_j\eta\gamma\theta=(v'_j\eta)\gamma\theta=
        v_j\gamma\theta\]
      because $\A'$ is $\Delta[\tau^+]$-more general than $\A$
      for $\eta$. Finally,
      \[v_j\gamma\theta=(v_j\gamma)\theta=v_j\theta\]
      because $Var(\gamma)\subseteq Var(c_1,c'_1)$ and
      $Var(c_1,c'_1)\cap Var(Q)=\emptyset$.
    \end{itemize}
    Consequently, we have proved that
    \[q(v'_1,\dots,v'_m)
      \text{ is $\Delta[\tau^+]$-more general than }
      q(v_1,\dots,v_m)\theta\text{ for }
      \sigma\eta\gamma\theta \;. \]
    As $q(v'_1,\dots,v'_m)$ denotes any atom of $\A'$, we have
    proved that
    \begin{equation}\label{lem-propagation-eq3}
      \A' \text{ is $\Delta[\tau^+]$-more general than } \A\theta
      \text{ for } \sigma\eta\gamma\theta\;.
    \end{equation}
  \item Let $A':=q(v'_1,\dots,v'_m) \in \B'$. As
    $\B=\B'\gamma$, then $\B'$ and $\B$ have the same length.
    Moreover, if $k$ denotes the position of $A'$ in $\B'$,
    then the $k^{\text{th}}$ atom of $\B$ has form
    $q(v_1,\dots,v_m)$.
    \begin{itemize}
    \item For each $\langle j\mapsto u_j\rangle \in \tau^+(q)$,
      $v'_j$ is an instance of $u_j$ (because $\tau^+$ is DN for
      $c'_1$ and \textbf{(DN4)} in Definition~\ref{def-DN}).
    \item For each $j \in [1,m]\setminus Dom(\tau^+(q))$ we have:
      \[v'_j\sigma\eta\gamma\theta=(v'_j\sigma)\eta\gamma\theta
        =v'_j\eta\gamma\theta\]
      because, by \textbf{(DN3)} in Definition~\ref{def-DN},
      \[Var(v'_j) \cap
      Var\big(\{s'_i \; | \; i \in Dom(\tau^+(p))\}\big)=\emptyset\]
      with
      $Dom(\sigma)=Var\big(\{s'_i \; | \; i \in Dom(\tau^+(p))\}\big)$.
      Moreover,
      \[v'_j\eta\gamma\theta=(v'_j\eta)\gamma\theta=v'_j\gamma\theta\]
      because $Var(\eta)\subseteq Var(Q,Q')$ and
      $Var(c'_1)\cap Var(Q,Q')=\emptyset$. Finally,
      \[v'_j\gamma\theta=(v'_j\gamma)\theta=v_j\theta\]
      because $\B=\B'\gamma$.
    \end{itemize}
    Consequently, we have proved that
    \[q(v'_1,\dots,v'_m)
      \text{ is $\Delta[\tau^+]$-more general than }
      q(v_1,\dots,v_m)\theta \text{ for } \sigma\eta\gamma\theta\;.\]
    As $q(v'_1,\dots,v'_m)$ denotes any atom of $\B'$, we have
    established that
    \begin{equation}\label{lem-propagation-eq2}
      \B' \text{ is $\Delta[\tau^+]$-more general than } \B\theta
      \text{ for } \sigma\eta\gamma\theta\;.
    \end{equation}
  \item Finally, by the same reasoning as for $\A$ and $\A'$
    above, we show that
    \begin{equation}\label{lem-propagation-eq3-1}
      \C' \text{ is $\Delta[\tau^+]$-more general than }
      \C\theta \text{ for } \sigma\eta\gamma\theta\;.
    \end{equation}
  \end{itemize}
  So, we conclude from~(\ref{lem-propagation-eq3}),
  (\ref{lem-propagation-eq2}) and (\ref{lem-propagation-eq3-1})
  that $(\A',\B',\C')$ is $\Delta[\tau^+]$-more general than
  $(\A,\B,\C)\theta$ for $\sigma\eta\gamma\theta$
  \ie{} that
  \begin{equation}\label{lem-propagation-eq3bis}
    (\A',\B',\C') \text{ is $\Delta[\tau^+]$-more general than }
    Q_1 \text{ for } \sigma\eta\gamma\theta\;.
  \end{equation}
  But, by the Technical Lemma~\ref{technical-lemma},
  $\sigma\eta\gamma\theta$ is a unifier of $p(s'_1,\dots,s'_n)$
  and $p(t'_1,\dots,t'_n)$. As $\theta'$ is an mgu of
  $p(s'_1,\dots,s'_n)$ and $p(t'_1,\dots,t'_n)$ (because
  $Q'\lral^{\theta'}_c Q'_1$ with $c'_1$ as input clause), there
  exists $\delta$ such that $\sigma\eta\gamma\theta=\theta'\delta$.
  Therefore, we conclude from~(\ref{lem-propagation-eq3bis})
  that $(\A',\B',\C')$ is $\Delta[\tau^+]$-more general than
  $Q_1$ for $\theta'\delta$. But this result implies that
  $(\A',\B',\C')\theta'$ is $\Delta[\tau^+]$-more
  general than $Q_1$ for $\delta$ \ie{} that
  $Q'_1$ is $\Delta[\tau^+]$-more general than $Q_1$ for
  $\delta$. Finally, we have proved that
  $Q'_1$ is $\Delta[\tau^+]$-more general than $Q_1$.\qed
\end{proof}

Using the Propagation
Lemma~\ref{lemma-classical-propagation},
the preceding result can be extended as follows.
\begin{proposition}[$\Delta$-Propagation]
  \label{proposition-propagation}
  If there exists an SLD-derivation step
  $Q' \lral^{\theta'}_c Q'_1$
  where $p(t'_1,\dots,t'_n)$
  is the selected atom then $Q'_1$ is
  $\Delta[\tau^+]$-more general than $Q_1$.
\end{proposition}
\begin{proof}
  Let $c'_1$ be the input clause used in
  $Q' \lral^{\theta'}_c Q'_1$.
  Take a variant $Q''$ of $Q$ such that
  \[Var(Q'') \cap Var(c'_1) = \emptyset\]
  and a variant $c''_1$ of $c$ such that
  \[Var(c''_1) \cap Var(Q'') = \emptyset \; .\]
  Then, the SLD-resolvent $Q''_1$ of $Q''$ and $c$ exists
  with the input clause $c''_1$ and with the atom selected
  in the same position as in $Q$.
  So, for some
  $\theta''$, we have $Q'' \lral^{\theta''}_c Q''_1$
  with input clause $c''_1$.
  Consequently, we have:
  \[Q \lral^{\theta}_c Q_1 \quad \text{and} \quad
    Q'' \lral^{\theta''}_c Q''_1 \; .\]
  $Q$ and $Q''$ are instances of each other because
  $Q''$ is a variant of $Q$.
  So, by the Propagation
  Lemma~\ref{lemma-classical-propagation}
  used twice, $Q''_1$ is an instance of $Q_1$
  and $Q_1$ is an instance of $Q''_1$. So, 
  \begin{equation}\label{lem-propagation-eq4}
    Q''_1 \text{ is a variant of } Q_1 \; .
  \end{equation}
  But we also have
  \[Q'' \lral^{\theta''}_c Q''_1 \quad \text{and} \quad
    Q' \lral^{\theta'}_c Q'_1\]
  with input clauses $c''_1$ and $c'_1$, with $Q'$ that
  is $\Delta[\tau^+]$-more general than $Q''$
  (because $Q''$ is a variant of
  $Q$ and $Q'$ is $\Delta[\tau^+]$-more general than $Q$)
  and $Var(Q'') \cap Var(c'_1) = \emptyset$.
  So, by Lemma~\ref{lemma-propagation},
  \begin{equation}\label{lem-propagation-eq5}
    Q'_1 \text{ is $\Delta[\tau^+]$-more general than }
    Q''_1 \;.
  \end{equation}
  Finally, from~(\ref{lem-propagation-eq4})
  and~(\ref{lem-propagation-eq5}) we have:
  $Q'_1$ is $\Delta[\tau^+]$-more general than $Q_1$.
  \qed
\end{proof}

\subsection{Epilogue}
Theorem~\ref{theorem-DN-associated} is a direct consequence
of the following result.

\begin{proposition}[One Step $\Delta$-Lifting]
  Let $c'$ be a variant of $c$ variable disjoint with $Q'$.
  Then, for some $\theta'$ and $Q'_1$,
  \begin{itemize}
  \item $Q' \lral^{\theta'}_c Q'_1$ where $c'$ is the input
    clause used,
  \item $Q' \lral^{\theta'}_c Q'_1$ is a
    $\Delta[\tau^+]$-lift of $Q \lral^{\theta}_c Q_1$.
  \end{itemize}
\end{proposition}
\begin{proof}
  Let $c'_1:=p(s'_1,\dots,s'_n)\leftarrow \B'$ be a
  variant of $c_1$.
  Then there exists a renaming $\gamma$ such that
  $Var(\gamma) \subseteq Var(c_1,c'_1)$ and
  $c_1 = c'_1\gamma$.
  Suppose also that
  \[Var(c'_1) \cap Var(Q,Q') = \emptyset \; .\]
  By the Technical Lemma~\ref{technical-lemma},
  $p(s'_1,\dots,s'_n)$ and $p(t'_1,\dots,t'_n)$ unify.
  Moreover, as $Var(c'_1) \cap Var(Q') = \emptyset$,
  $p(s'_1,\dots,s'_n)$ and $p(t'_1,\dots,t'_n)$ are
  variable disjoint. Notice that the following claim holds.
  \begin{claim}
    Suppose that the atoms $A$ and $H$ are variable disjoint
    and unify. Then, $A$ also unifies with any variant $H'$
    of $H$ variable disjoint with $A$.
  \end{claim}
  \begin{proof}
    For some $\gamma$ such that $Dom(\gamma)\subseteq Var(H')$,
    we have $H=H'\gamma$. Let $\theta$ be a unifier of $A$
    and $H$. Then,
    $A\gamma\theta=A\theta=H\theta=H'\gamma\theta$,
    so $A$ and $H'$ unify.
    \qed
  \end{proof}

  Consequently, as $c'$ is a variant of $c'_1$ and
  $c'$ is variable disjoint with $Q'$,
  $p(t'_1,\dots,t'_n)$ and the head of $c'$
  unify. As they also are variable disjoint, we have
  \[Q' \lral^{\theta'}_c Q'_1\]
  for some $\theta'$ and $Q'_1$ where $p(t'_1,\dots,t'_n)$
  is the selected atom and $c'$ is the input clause
  used. Moreover, by the $\Delta$-Propagation
  Proposition~\ref{proposition-propagation},
  $Q'_1$ is $\Delta[\tau^+]$-more general than $Q_1$.
  \qed
\end{proof}

\end{document}